# Automating Analysis of Neutron Scattering Time-of-Flight Single Crystal Phonon Data

**Dmitry Reznik** [1,2,*] **and Irada Ahmadova** [1,2]

[1] Department of Physics, University of Colorado-Boulder, Boulder, CO 80309, USA;
[2] Center for Experiments on Quantum Materials, University of Colorado-Boulder, Boulder, CO 80309, USA;
* Correspondence: dmitry.reznik@colorado.edu

**Abstract:** This article introduces software called Phonon Explorer that implements a data mining workflow for large datasets of the neutron scattering function, $S(\mathbf{Q}, \omega)$, measured on time-of-flight neutron spectrometers. This systematic approach takes advantage of all useful data contained in the dataset. It includes finding Brillouin zones where specific phonons have the highest scattering intensity, background subtraction, combining statistics in multiple Brillouin zones, and separating closely spaced phonon peaks. Using the software reduces the time needed to determine phonon dispersions, linewidths, and eigenvectors by more than an order of magnitude.

## 1. Introduction

Many scientific studies require comprehensive investigations of phonon dispersions and other lattice dynamical effects in crystalline solids. For example, the phonon spectrum determines in large part the thermal conductivity of materials [1]; phonons have been implicated in charge density wave formation and structural instabilities [2–4]; so-called phonon anomalies may indicate the Fermi surface nesting or a coupling to the superconducting gap [5,6]; comparing experimental phonon dispersions with model predictions can validate or invalidate theoretical models [7].

One way to obtain phonon spectra is from density functional theory (DFT) calculations [8]. These are getting increasingly sophisticated and accurate. However, they still often fall short, especially in materials where strong electronic correlations and magnetoelastic coupling are present. For example, recent work on a high temperature superconductor $HgBa_2CuO_4$ showed that different commonly used functionals and approximations within the DFT framework give different phonon energies [9]. DFT results for phonon eigenvectors are often only qualitatively accurate even when dispersions closely match experiment, especially in the vicinity of branch crossings. Furthermore, these calculations are difficult or impossible for materials with very large unit cells. Moreover, sometimes the purpose of scientific investigations is to identify reciprocal space points where phonon energies and/or linewidths deviate from DFT predictions [4]. Therefore, it is important to be able to extract phonon energies, linewidths, and eigenvectors directly from the data without relying on DFT.

Inelastic neutron or x-ray scattering experiments are powerful tools for measuring phonon dispersions and, to a lesser extent, linewidths and eigenvectors [10]. Due to recent advances in neutron scattering instrumentation, especially the time-of-flight (TOF) instruments, scientists can now routinely measure complete scattering spectra of single crystal samples covering a large range of momenta. Experiments measuring phonon spectra are performed using a monochromatic neutron beam with the full width at half maximum (FWHM) experimental energy resolution on the order of 3–5% of the incident energy, $E_i$. Typically, 3–5 meV resolution is sufficient, so the largest $E_i$ that is used can be on the order of 100 meV. ARCS spectrometer at the Spallation Neutron Source (SNS) [11] can detect neutrons scattered as much as 135° away from the incident beam direction and as little as 3°. This instrument design allows measurement of the scattering intensity, $S(\mathbf{Q},\omega)$ [10], over a range of momentum transfers to the neutron $\mathbf{Q}$ whose magnitude ranges from nearly zero to $\approx 10\text{Å}^{-1}$. Here $\hbar\omega$ is the energy transfer to the neutron.

For a typical material with complex structure whose unit cell is around 4 Å$^{-1}$, the first Brillouin zone (BZ) size is on the order of less than 2Å$^{-1}$. For many layered materials one of the unit cell dimension can be 10–30Å, which means that one of the dimensions of the BZ is less than 1Å$^{-1}$. These order of magnitude estimates mean that the volume of one BZ is on the order of 5(Å$^{-1}$)$^3$. In contrast, the range of $\mathbf{Q}$s for which $S(\mathbf{Q}, \omega)$ is measured is on the order of 10 × 10 × 10 (Å$^{-1}$)$^3$ = 1000(Å$^{-1}$)$^3$. So the four-dimensional $S(\mathbf{Q}, \omega)$ often covers hundreds, sometimes thousands BZs.

These datasets have sufficient coverage for reconstructing phonon dispersions and eigenvectors as well as the linewidths that are comparable to or larger than the instrument resolution. The comprehensive datasets can also be searched for interesting unexpected effects.

If the goal of the data analysis is to extract phonon dispersions and linewidths in energy, one needs to perform a one-dimensional cut for each $\mathbf{Q}$ or interest: Fix $\mathbf{Q} = \mathbf{Q}_0$ and look at $S(\mathbf{Q}, \omega)$ as a function of $\omega$. Considering the large volume of $\mathbf{Q}$-space covered, and the necessity to look at many Brillouin zones, a comprehensive coverage of reciprocal space without significant automation is not practical. Even limiting the investigation to high-symmetry directions in momentum space still requires performing thousands of cuts if one wants to look for peaks corresponding to every phonon in every BZ. This is extremely time-consuming, even with the aid of ad-hoc scripts, and is usually not done. As a result, most of the data that could potentially contain useful information are never looked at. Lack of an established workflow and shortcomings in available software confine researchers to small ranges of momentum transfer that they see as most promising.

This practice is justified on a certain level, since the information in multiple BZs is partially redundant: According to Bloch's theorem, the good momentum quantum number in crystals is crystal momentum denoted here by $\mathbf{q}$, not total momentum $\mathbf{Q}$ [12], so all excitations can in principle be measured when $\mathbf{Q}$ is confined to a single BZ.

However, there are significant problems with this approach. First, the neutron scattering intensity strongly depends on BZ [10], and unexpected interesting features contained in the data may be missed if only a subset of BZs is looked at. Second, it is possible to mistake several overlapping resolution-limited peaks for one broad peak. Third, accurate eigenvector determination requires a systematic examination of multiple BZs. Fourth, energy-dependent features in the background can result in incorrect peak assignments. Finally, valuable experimental time may be wasted to accumulate statistics in each BZ instead of combining data from multiple zones.

This article introduces a recently released software tool, Phonon Explorer, that implements a data analysis workflow that solves these problems and includes automated background subtraction.

The workflow includes the following steps:

1. Finding relevant Brillouin zones
2. Background determination and subtraction
3. Optimization of binning
4. Extracting phonon dispersions, linewidths, and eigenvectors by multizone fit.

It has been successfully applied to cases where a measurement of all phonons is desired as well as to projects focusing on specific phonons or phonon branches [9,13,14]. Here we focus on phonons mostly because their scattering intensity is greatest at large wavevectors where manually going through every Brillouin zone is especially time-consuming. Many parts of this discussion can apply equally well to investigations of magnetic and other nonphononic excitations although the current version of the software does include such use cases.

We now go through each step of the workflow one-by-one and present several examples from recently completed or in-progress research projects. At the end, we describe the main features of Phonon Explorer software used to generate the examples shown in the figures while leaving the detailed description to the user manual posted online.

## 2. Determination of Relevant Brillouin Zones

As already mentioned above, according to Bloch's theorem, every phonon in a periodic crystal is characterized by a reduced wavevector, **q** that uniquely defines crystal momentum [12]. Therefore S(**Q**, ω), should have a peak in every Brillouin zone (BZ) corresponding to each phonon. To maximize the use of the data, the analysis should cover every BZ. The scattering intensity of each phonon is determined by the phonon structure factors, which depend on the BZ and derive from the phonon eigenvectors. In most cases, especially in materials with many atoms in the unit cell, the scattering intensity of each phonon is appreciable only in a small fraction of the Brillouin zones.

We illustrate this point by using an S(**Q**, ω) dataset measured on the MERLIN spectrometer at ISIS [15] on $Ba_8Ga_{16}Ge_{30}$ [16] with the incident neutron energy of 52 meV. This material belongs to the clathrate type-I structure with a host cage framework of Ga and Ge atoms holding Ba guest atoms inside the cages. We are currently analyzing neutron scattering data with the purpose of understanding the small phonon thermal conductivity of this material. Here we would like to show how the framework proposed here can be used to identify phonon modes at the zone center between 5 and 15 meV. We expect several closely spaced modes whose energy separation is significantly smaller than the energy resolution of the instrument.

The first step of the workflow is to identify zones where scattering from phonons of interest has significant intensity. The dataset covers nearly 1500 Brillouin zones. We performed constant **Q** cuts at the zone center of every zone and obtained 1454 cuts. Figure 1a shows 10 typical cuts. Spectra plotted in blue are nearly flat and do not show any identifiable one-phonon scattering. The red spectra show mostly broad peaks originating from one or several overlapping phonon peaks.

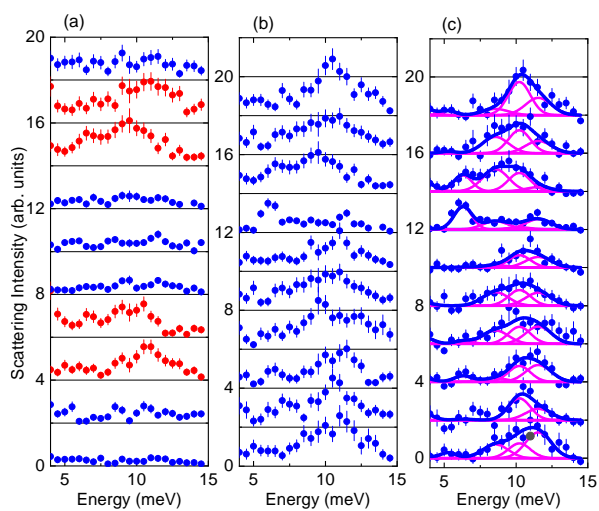

**Figure 1.** Zone center phonon data obtained from measured S (**Q**, ω) of $Ba_8Ga_{16}Ge_{30}$. The binning was ΔH = ΔK = ΔL = ±0.1 reciprocal lattice units (r.l.u.). The wavevectors (H,K,L) in order from top to bottom are: (**a**) **Q** = (2,−1,−3), (−2,−1,−7), (−2,−1,−9), (−4,0,−11), (−4,0,−3), (−4,0,−4), (−4,0,−6), (−1,0,−10), (−10,−9,2), (−7,−9,3); For (b,c): **Q** = (−9,−10,−4), (−8,−7,−10), (−7,−8,−9), (−1,−4,−3), (−10,−9,2), (−3,−4,−9), (−5,−9,−8), (−6,−8,−9), (−7,−10,−2), (−1,0,−10). (**a**,**b**) data points represent raw data. (**c**) data points represent background-subtracted data. Magenta curves represent individual phonon peaks determined through the multizone fit. Blue curves are fit results. Peak positions based on the fit are: 5.3 ± 0.1, 6.3 ± 0.1 8.6 ± 0.15, 10.2 ± 0.1, and 11.6 ± 0.2 meV.

Then we visually examined all 1454 cuts and identified 35 BZs where the phonon signal similar to the one observed in the red spectra is present. Phonon Explorer software lays out the plots in such a way that it actually took about 30 min. Figure 1b shows ten of these cuts chosen at random.

## 3. Background Determination and Subtraction

Once the useful Brillouin zones are identified, background must be subtracted from the data. Correctly determining the background is challenging, because it is impossible to separate neutrons scattered by phonons of the investigated sample from neutrons that scatter into the same pixel via a different process. For example, background can originate from phonons of the sample holder or multiple scattering.

Phonon Explorer software supports several possible background determinations. The simplest one subtracts a constant to make the minimum of the data close to zero. This method may work if the energy range of the dataset is small and one can neglect the energy-dependence of the background. One needs to be sure that there are no significant features originating from the sample holder and the phonon signal is much stronger than the background. Data points in Figure 1c show the cuts in Figure 1b after subtraction of constant background. In addition to automatic constant background subtraction, Phonon Explorer includes an option to make further manual adjustments by subtracting user-defined linear background.

We now discuss a more sophisticated background determination procedure needed for data with small signal-to-background ratio where features in the background can make a significant contribution to phonon peaks.

If we are interested in the phonon spectrum at a specific **Q**, nearby wavevectors **Q'** can often be found where the phonon of interest is at a different energy due to dispersion or the phonon structure factor vanishes. If $|\mathbf{Q}| = |\mathbf{Q'}|$ then the spectrum of the polycrystalline sample holder or incoherent scattering, which depends only of the wavevector magnitude, is unchanged. Then the background can be read off as a point-by-point minimum of the cuts at **Q** and **Q'**.

Alternatively, one may choose the **Q'** vectors at random and then the background value at each energy would represent the minimum of the intensities of the points. This procedure was followed in Ref. [17], except the condition $|\mathbf{Q}| = |\mathbf{Q'}|$ was not enforced. Instead the intensity of each spectrum was multiplied by $|\mathbf{Q'}|^2$ based on the assumption that most background intensity comes from phonon scattering whose intensity is generally proportional to $|\mathbf{Q}|^2$ [10]. The implicit assumption is that one-phonon scattering from the sample will vanish at every energy at one of the **Q'**s.

If the data are noisy, simply taking a point-by point minimum of the data at different **Q**'s will underestimate the background. In this case it is more accurate to make the background equal to the point-by point minimum of smoothed data rather than raw data. Using a standard smoothing procedure such as taking the average of adjacent points or doing an interpolation produces features that originate from noise. We found that fitting to multiple Gaussian peaks whose initial spacing and minimal width are on the order of the instrument resolution is a better way to smooth data. Some tuning of the peak spacing as well as the lower bound for the peak width is typically required to obtain the correct smooth curve.

Figure 2 illustrates how the background was determined in a recent investigation of Cu-O bond stretching phonons where spectrum was measured on a small sample of $HgBa_2CuO_{4+\delta}$ where strong background originates from the sample holder as well as from incoherent scattering by the sample. We want to subtract background from the 69 meV phonon at **Q** = (5,0,1.5). The raw data are shown in Figure 2a.

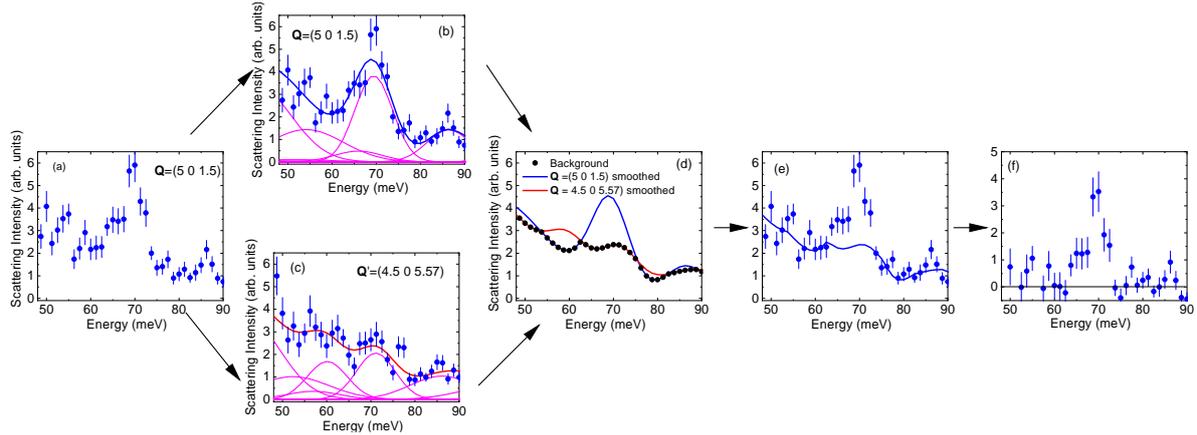

**Figure 2.** Background subtraction workflow described in the text. (**a**) Raw data; (**b**,**c**) Raw data with a smooth curve represented by the blue line. Blue line is obtained from fitting to a sum of Gaussian peaks represented by magenta lines; (**d**) Background is the point-by-point minimum of the two smooth curves; (**e**) Raw data with background; (**f**) Background-subtracted data.

It is known that the bond-stretching branch disperses downward toward the zone boundary, and therefore we do not expect any phonons near this energy at H = 4.5. In order to keep the magnitude of the wavevector the same, we increase L to 5.57, so that **Q′** = (4.5, 0, 5.57). The cut at **Q′** is shown in Figure 2c.

The data at both **Q** and **Q′** (Figure 2b,c) are fit to a sum of several Gaussian peaks so that a smooth curve goes through the data as described above. The results of these fits represented by the blue and red curves are plotted in Figure 2c. The background represented by the black dots is the point-by-point minimum of these curves. The red curve goes above the blue curve around 60 meV, which is the expected position of the zone boundary bond-stretching phonon. Due to the layered structure, it should only weakly depend on L.

Figure 2e shows the raw data at **Q** together with the background, and the background-subtracted spectrum is shown in Figure 2f.

The software supports extending this procedure to use several **Q′** vectors to make sure one-phonon scattering is not picked up by the background.

This procedure does not separate background due to multiple scattering from the rest of the data and other contributions to the background. Sometimes it is possible to calculate multiple scattering if the phonon spectrum and sample shape are known. In this case, we recommend subtracting it from the data first. However, there are many possible contributions to the background whose intensity is extremely difficult to estimate. For example, in some cases multiple scattering from the sample and the sample environment has been shown to introduce a broad peak at 17 meV [18]. Its intensity should not strongly depend on the sample rotation angle, so it would be subtracted following the procedure described here.

It is important to keep in mind that there can be some energies at which one-phonon scattering from the sample at nearby wavevectors is always present, so the background determined based on the data at **Q′**'s will inevitably contain one-phonon intensity that we do not want to subtract. In this case, it is better to subtract linear background (see above). It is the least accurate method, because background can have features, but in many cases it is the only approach that can be taken.

## 4. Refinement of Binning

As the data analysis progresses, it is necessary to reassess whether the originally chosen binning in crystal momentum is optimal. Binning that is too fine, will inevitably degrade statistics, whereas binning that is too coarse will degrade momentum space resolution. Optimal binning can also depend on wavevector: Coarser binning at high symmetry points such as the zone center and the zone boundary can give the same effective **Q**-resolution as a finer binning away from either reciprocal space point. Ideally one wants to repeat the steps in previous two sections for different binning values to see which works best.

One has to be careful about rehistogramming, i.e., rebinning data that have already been binned, which will give incorrect intensities and error bars. To speed up data analysis we recommend to ignore this problem initially. However, once the optimal binning has been identified, the raw data should be histogrammed starting from the event mode data using this optimal binning and the analysis should be repeated.

**5. Multizone Fit**

According to Bloch's theorem, peak positions and linewidths of phonons at a given reduced wavevector are the same in different BZs, but amplitudes vary from zone to zone following the phonon structure factor. Mulitizone fitting takes advantage of this property by fitting constant Q cuts in different Brillouin zones at the same reduced wavevector simultaneously. The first step is to guess the number and positions of the peaks based on visual inspection of the data. Then, multizone fit optimizes the peak positions, linewidths, and amplitudes while constraining positions and linewidths of the same phonon to be the same in different zones. If fit results give poor agreement with the data, the number of peaks needs to increased and the fit repeated. Using data from more than one BZ automatically differentiates phonon peaks in the constant-Q scans from spurious peaks such as multiple scattering peaks, since the former obey Bloch's theorem and the latter do not. Multizone fitting greatly improves the precision of the peak position and linewidth determination, as it effectively combines statistics from different Brillouin zones.

Typically, the fit function should be a sum of Lorentzian peaks convoluted with the instrument resolution function. The resolution function of TOF instruments has an asymmetric functional form. Sometimes the phonon lineshape is not a Lorentzian, and special care needs to be taken in such cases. The current version of the Phonon Explorer software uses Gaussian peak shapes to increase the speed. These do not reproduce asymmetry of the resolution-limited measured phonon peaks. Aside from that, they give a good agreement with the data. The option to use a more accurate functional form of the peaks will be included in future releases.

Figure 1c shows how we use multizone fitting to determine peak positions of poorly resolved phonons. Note that the actual fit was based on 35 different wavevectors, not just the 10 shown in the figure.

Figure 3 illustrates the multizone fit for the bond-stretching phonon in Hg1201 at 69 meV. We use data from seven Brillouin zones, and the dependence of the phonon scattering intensity from zone to zone allows us to determine the eigenvector of this phonon. (See Ref. [9] for more details)

In Figure 4 Multizone fit is used to determine the dispersion of the bond-stretching phonons and investigate their possible anticrossing with the bond-bending branch. Note that the effective energy resolution depends on binning and the slope of the phonon dispersion. Based on the relative intensities of the upper and lower branches in different zones it was found that an anticrossing does not occur.

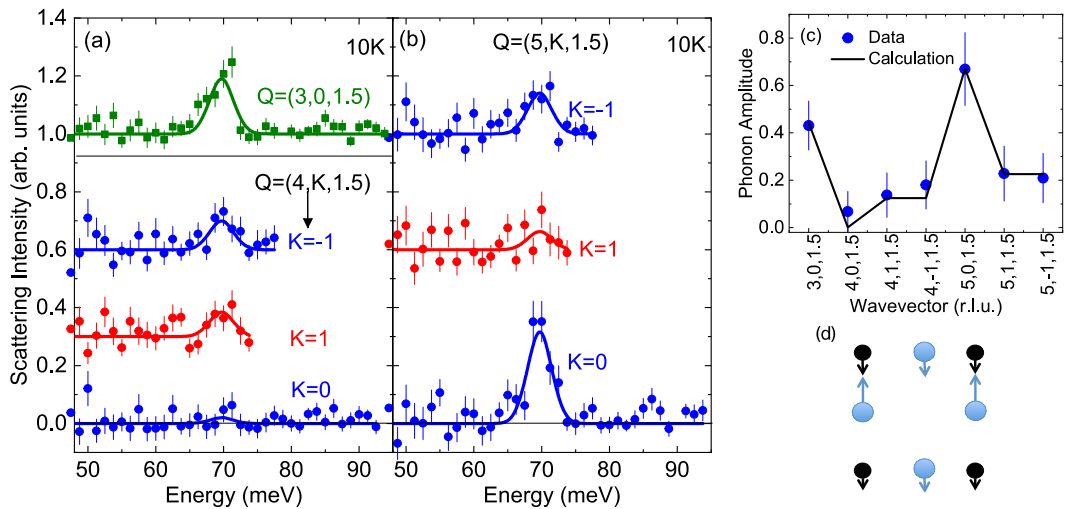

**Figure 3.** Zone-center bond-stretching phonon. (**a**,**b**) Solid line is the result of a multizone fit (see text) to the background-subtracted data (symbols). (**c**) Integrated intensity of the phonon at different wavevectors compared with the prediction based on the eigenvector shown in (**d**). (d) Eigenvector of the bond-stretching phonon. Small black circles represent Cu ions, large blue circles represent O. Arrows represent the phonon eigenvector drawn to scale. The eigenvector was extracted from phonon intensities in (a,b). (From Ref. [9]).

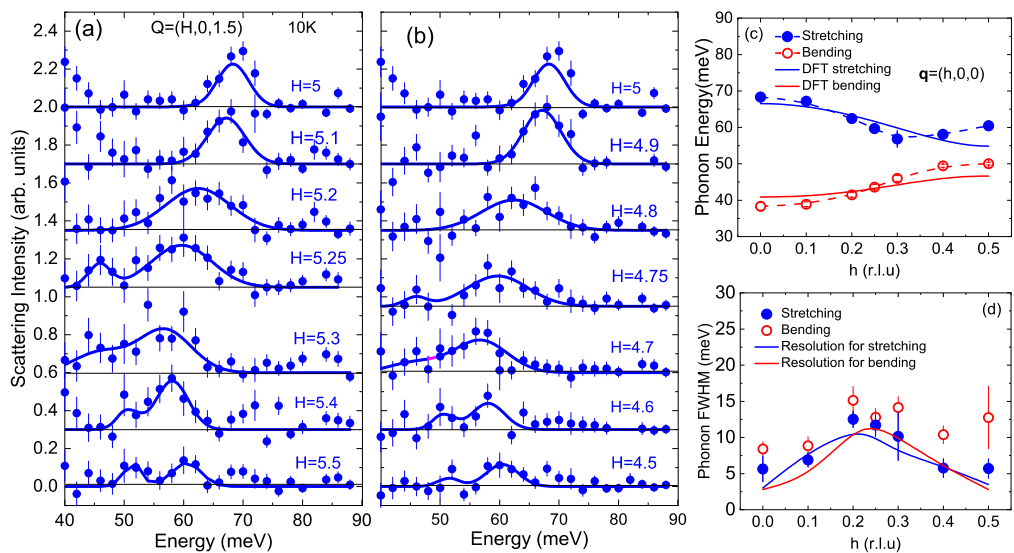

**Figure 4.** Cu-O bond-stretching and bond-bending longitudinal phonon dispersion along [100]. (**a**,**b**) Background-subtracted data at wavevectors with large bond-stretching phonon structure factor and small bond-bending structure factor. Binning: $\Delta H = \pm 0.07$, $\Delta K = \pm 0.08$, and $\Delta L = \pm 3.5$. (**c**) Dispersions of the bond-stretching (blue) and bond-bending phonon branches (red). For the bond-stretching phonon peaks the multizone fit was made to the data at each **q** in the two zones shown in (a,b). The bond-bending (lower) branch peak positions were fixed to the values obtained based on wavevectors adjacent to (4,0,0), (4,1,0), and (4,−1,0). (**d**) Raw-data peak widths (FWHM) extracted from the same fits as in (c). The effective resolution width increases when the dispersion is steeper. (From Ref. [9]).

## 6. Brief description of Phonon Explorer Software.

The data analysis used to illustrate the workflow presented in the figures above has been performed entirely using the Phonon Explorer software [19] that implements most of the workflow outlined above. The software is written in Python and can extract constant **Q** slices by calling lower level routines in the software packages that perform histogramming and taking 1D or 2D slices of the 4D datasets. Current version of the Phonon Explorer is compatible with one such package developed at the Spallation Neutron Source in the USA called MANTID [20,21] as well as another package developed at ISIS in the UK called HORACE [22]. To use HORACE, which is written in Matlab, the software also includes a short Matlab routine that interfaces with HORACE. The code is 100% object-oriented and scalable.

The software allows users to investigate S (**q**, ω) at a specific reduced wavevector **q** working with all total wavevectors, **Q**, that correspond to this **q** at once. Alternatively, one can enter a list of total wavevectors of interest into a text file and work only with these wavevectors. The software also encourages a new conceptual way of thinking, that is in line with the full capabilities of the TOF technique.

Automation of the workflow leads to an order of magnitude increase of the speed with which the data are processed making studies comparing phonon dispersions with DFT calculation results rapid and routine. Figure 5 shows phonon peak positions obtained with the help of Phonon Explorer compared with DFT results. Data points represent energies of the phonons with nonzero structure factor contained in the TOF dataset originally obtained to investigate magnetic excitations. Following the software-backed workflow discussed in this article ensured that all phonons that have a detectable intensity in the dataset were extracted and fit with minimal time and effort. Note good agreement between the calculation and the data except for the lowest branch at the Y-point. Such full cycle of comprehensive data analysis takes 2–3 weeks depending on the level of familiarity with the software.

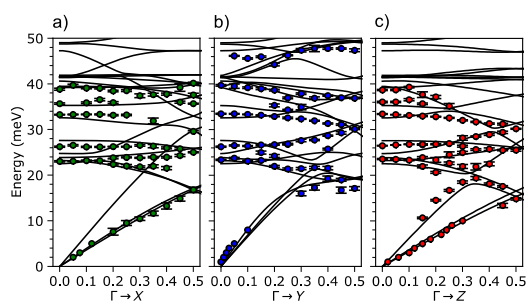

**Figure 5.** Phonon energies in FeP extracted from TOF data with the help of the Phonon Explorer software compared with DFT calculation results (**a**) Γ-X direction; (**b**) Γ-Y direction; (**c**) Γ-Z direction; (From Ref. [14]).

## 7. Conclusions

To summarize, Phonon Explorer software has been successfully used in three completed research projects spanning a variety of experimental conditions [9,13,14].

This software:

1. Helps to distinguish between a broad peak and multiple closely-spaced peaks (Figure 1).
2. Helps to distinguish between branch crossings and electron-phonon anomalies (Figure 4).
3. Improves accuracy of fitting data. (Figures 1–5)
4. Enables efficient search for new physics and data mining in TOF datasets (Figure 5).
5. Subtracts background from the data (Figure 2)
6. Automates full cycle data analysis (Figures 1–5)

We recommend installing the software on chopper spectrometers to give users the opportunity to perform rapid sophisticated real time data analysis in order to adjust experimental conditions and strategy during experiments.

A. Sokolik helped with interfacing the Python code with the HORACE package.

**Author Contributions:** D.R. Conceived the project, wrote the current Python version of the software, wrote the paper. I.A. Coded the prototype of the software

**Funding:**   This research was funded by the U.S. Department of Energy, Office of Basic Energy Sciences, Office of Science, under Contract No. DE-SC0006939

**Conflicts of Interest**: The authors declare no conflict of interest.